\documentclass[aps,twocolumn,floatfix,nofootinbib,nopreprintnumbers]{revtex4}
\usepackage{dcolumn}
\usepackage{multirow}
\usepackage{graphicx}
\usepackage{amssymb}
\usepackage{bm}
\usepackage{hyperref}
\usepackage{epstopdf}
\usepackage{color}
\usepackage{mathrsfs}
\usepackage{amsmath,amssymb,amsthm}
\usepackage{rotating}
\usepackage{sverb, longtable}
\usepackage{subfigure}

\usepackage[graphicx]{realboxes}
\usepackage{adjustbox}
\begin{document}
	
\newcommand{\DW}[1]{\textcolor{blue}{{DW:[\bf#1]}}}	

\title{Do low-redshift observations \emph{open} the doors to an \emph{open} universe?}

\author{Deng Wang$^1$}
\email{dengwang@ific.uv.es}
\author{Olga Mena$^1$}
\email{omena@ific.uv.es}
\author{Salvatore Capozziello$^{2,3,4}$}
\email{capozziello@na.infn.it}
\author{David Mota$^5$}
\email{d.f.mota@astro.uio.no}

\affiliation{
	$^1${Instituto de F\'{i}sica Corpuscular (CSIC-Universitat de Val\`{e}ncia), E-46980 Paterna, Spain} \\
    $^2${Dipartimento di Fisica “E. Pancini”, Università di Napoli “Federico II”, Via Cinthia Edificio 6, I-80126, Napoli, Italy}\\
    $^3${Scuola Superiore Meridionale, Largo San Marcellino 10, 80138, Naples, Italy}\\
	$^4${INFN, Sezione di Napoli, Complesso Universitario di Monte S. Angelo, Via Cinthia Edificio 6, 80126, Naples, Italy}\\
    $^5$\mbox{Institute of Theoretical Astrophysics, University of Oslo, P.O. Box 1029 Blindern, N-0315 Oslo, Norway}
    }

\begin{abstract}

The detection of a significant deviation from a zero curvature would have profound consequences for inflationary theories and fundamental physics. Relative to high-redshift Planck's CMB measurements, indicating a $\sim 2\sigma$ evidence for a closed universe, low-redshift observations of BAO and SN Ia have the advantages of weak dependence on early universe physics, independently observational systematics, and strong redshift dependence of distances in constraining the cosmic curvature. Using the integrated observations from DESI BAO and SN Ia, we find an unexpected $2\sigma$ evidence for an open universe, regardless of the SN Ia sample employed. When considering DESI, SN Ia and the acoustic scale $\theta_\star$ data, the preference for an open universe exceeds the $3\sigma$ level, reaching $5\sigma$ for the case of DESY5 Supernovae data. Therefore, low-redshift observations favor an open universe, and this preference persists even when alternative high-redshift priors are adopted. Our results point to the existence of an additional tension between high- and low-redshift observations, present also in non-flat models beyond the minimal $\Lambda$CDM scheme, thereby challenging the standard inflationary predictions.

\end{abstract}

\maketitle

\section{Introduction} 

The standard cosmology, $\Lambda$-cold dark matter ($\Lambda$CDM), which characterizes the physical phenomena across multiple scales at the background and perturbation levels through the cosmic history~\cite{,Weinberg:2013agg,Turner:2022gvw,MukhanovPFoC,DodelsonSchmidtMC}, has been established and confirmed by various observations such as cosmic microwave background (CMB)~\cite{Planck:2018vyg,AtacamaCosmologyTelescope:2025blo,SPT-3G:2022hvq,WMAP:2003elm,Planck:2013pxb}, abundance of light elements~\cite{Olive:1999ij,Fields:2019pfx,Allahverdi:2020bys,Schoneberg:2024ifp} baryon acoustic oscillations (BAO)~\cite{2dFGRS:2005yhx,Beutler:2011hx,SDSS:2005xqv,Blake:2011en,BOSS:2013rlg,BOSS:2016apd,eBOSS:2017cqx,eBOSS:2020yzd,DESI:2024uvr,DESI:2024lzq,DESI:2025zpo} and type Ia supernovae (SN)~\cite{SupernovaSearchTeam:1998fmf,SupernovaCosmologyProject:1998vns} during the past almost three decades. However, it confronts at least two insurmountable challenges, i.e., the coincidence problem and the persistent cosmological constant conundrum~\cite{Weinberg:1988cp,Carroll:2000fy,Zlatev:1998tr,CosmoVerseNetwork:2025alb}. The former states why present-day densities of dark energy (DE) and dark matter (DM) are of the same order, while the latter is related to the value of cosmological constant inferred from current observations, that is much smaller than that expected from quantum field theory. Meanwhile, $\Lambda$CDM confronts two primary cosmological tensions that have arisen from recent observations, namely, the Hubble constant ($H_0$) tension~\cite{DiValentino:2021izs,DiValentino:2020zio} and the matter fluctuation amplitude ($S_8$) discrepancy~\cite{DiValentino:2020vvd,DiValentino:2020srs,Abdalla:2022yfr}. Both of them reflect an inconsistency between low and high-redshift observations of our universe. More concretely, the $H_0$ tension arises because the indirectly derived $H_0$ value from the Planck-2018 CMB observations, assuming $\Lambda$CDM, is more than $4\,\sigma$ lower than the direct measurement of the present-day cosmic expansion rate from the Hubble Space Telescope (HST)~\cite{Riess:2021jrx,Breuval:2024lsv,H0DN:2025lyy}, while the $S_8$ tension reveals that the strength of matter clustering in the linear regime measured by various low/intermediate-redshift probes including weak gravitational lensing ~\cite{DES:2021bpo,DES:2021zxv,KiDS:2020suj,HSC:2018mrq}, galaxy clustering~\cite{DES:2021wwk,Heymans:2020gsg,Lange:2023khv,Amon:2022ycy,Troster:2019ean}, cluster counts~\cite{DES:2020ahh,Mantz:2014paa} redshift space distortions~\cite{eBOSS:2020yzd,Macaulay:2013swa}, CMB lensing~\cite{Planck:2018lbu} and recent DESI-Planck cross-correlations~\cite{Karim:2024luk} is lower than that indirectly derived by the Planck-2018 CMB data~\cite{Planck:2018vyg} under $\Lambda$CDM. 
In this debate, a main role could be played by cosmography \cite{Capozziello:2018jya}.
These puzzles and discrepancies highlight the challenges faced by $\Lambda$CDM in depicting the fundamental physics and reconciling early and late universe observations. 

A non-negligible fact is that the above issues emerges in a flat universe, i.e., cosmic curvature $\Omega_k=0$. In practice, the importance of curvature on cosmological physics is always underestimated. This does not mean that cosmic curvature can definitely help in resolving these issues, but it plays a crucial role in describing the composition, evolution and fate of the universe. Curvature depicting the geometry of the universe influences significantly the manifestation of various cosmic tensions and problems. When studying the cosmological physics, more attention should be paid to the cosmic curvature, due to the following reasons: (i) it affects the background dynamics of the universe in the Friedmann equations; 
(ii) it can influence the growth of structures in the universe by changing the matter power spectrum measured by large scale structure observations and angular power spectrum of the CMB; (iii) since inflation~\cite{Guth:1980zm,Linde:1981mu,Starobinsky:1980te,Albrecht:1982wi,Kazanas:1980tx,Sato:1981qmu} predicts a flat universe, any deviation from flatness could challenge or refine inflationary models, reshaping our understanding of the very early universe. Measurements of curvature provide constraints on the initial conditions and dynamics of inflation; (iv) curvature has a more profound impact on the very long-term evolution of the universe since the big bang epoch, while dark matter, shaping the universe's structure, has a relatively short duration, and the dark energy component, while driving the late-time cosmic acceleration, predominantly exerts its influence at redshifts $z\lesssim1$. More details about the effects of curvature on cosmological physics are shown in ~\cite{Yang:2022kho,Gariazzo:2024sil}.
  
Considering a non-flat $\Lambda$CDM universe, Planck CMB observations suggest a slight preference for a negative curvature with $\Omega_k=-0.0105\pm 0.0066$, which is a $\sim2\,\sigma$ hint of a closed universe~\cite{Handley:2019tkm}, see also Refs.~\cite{DiValentino:2020hov,DiValentino:2020srs}. We recall here that the curvature parameter $\Omega_k$ decreases exponentially with time during inflation, but grows only as a power law along the radiation and matter-dominated phases, and therefore the standard inflationary prediction has been that curvature should be unobservably small
today. Nevertheless, by fine-tuning parameters it is possible to find in the literature inflationary models that generate both open or closed universes~\cite{Sasaki:1993ha,Bucher:1994gb,Linde:1995xm,Yamauchi:2011qq,Garcia-Bellido:1997zep,Ratra:2017ezv,Cespedes:2020xpn}. As previously stated, the detection of a significant deviation from a zero curvature would have profound consequences for inflation theory and fundamental physics. The Planck constraint is
mainly determined by both the priors on the Hubble constant and the effect of lensing smoothing on the power spectra. Since Planck observations show a slight
preference for more lensing than expected in the base $\Lambda$CDM cosmology~\cite{Planck:2015fie}, and since positive curvature increases the amplitude
of the lensing signal, this preference also brings the spatial curvature towards negative values, see Ref.~\cite{Planck:2015fie}

\begin{figure*}
	\centering
	\includegraphics[scale=0.5]{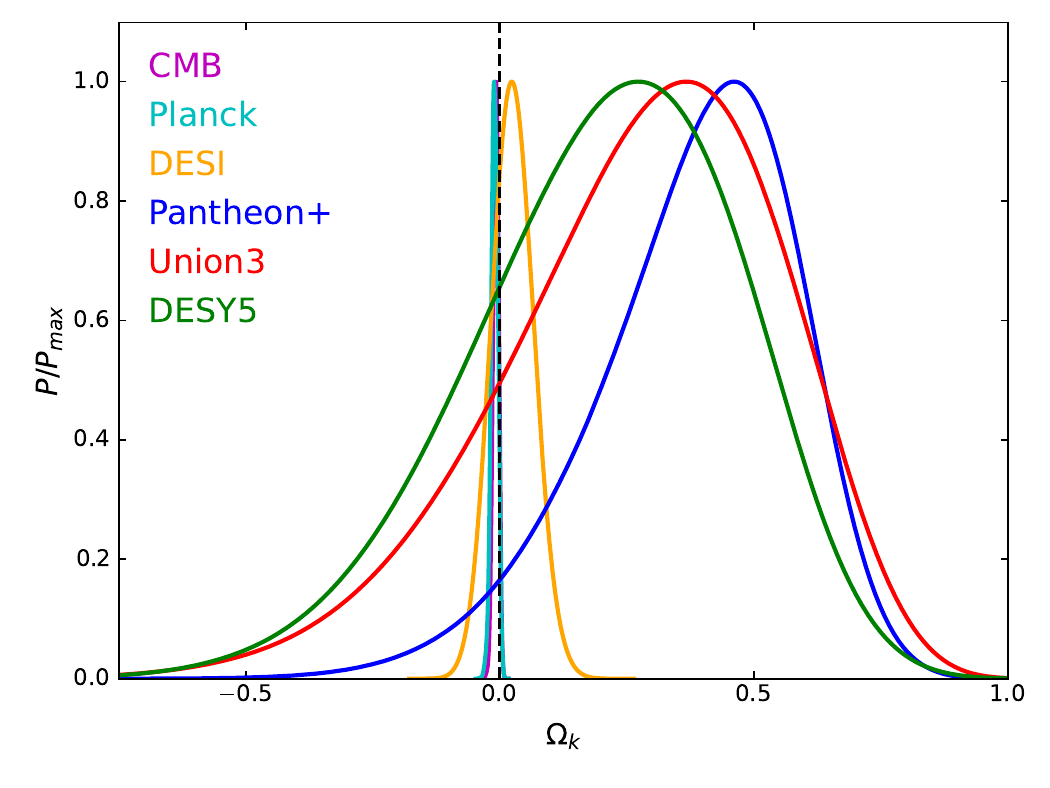}
	\includegraphics[scale=0.5]{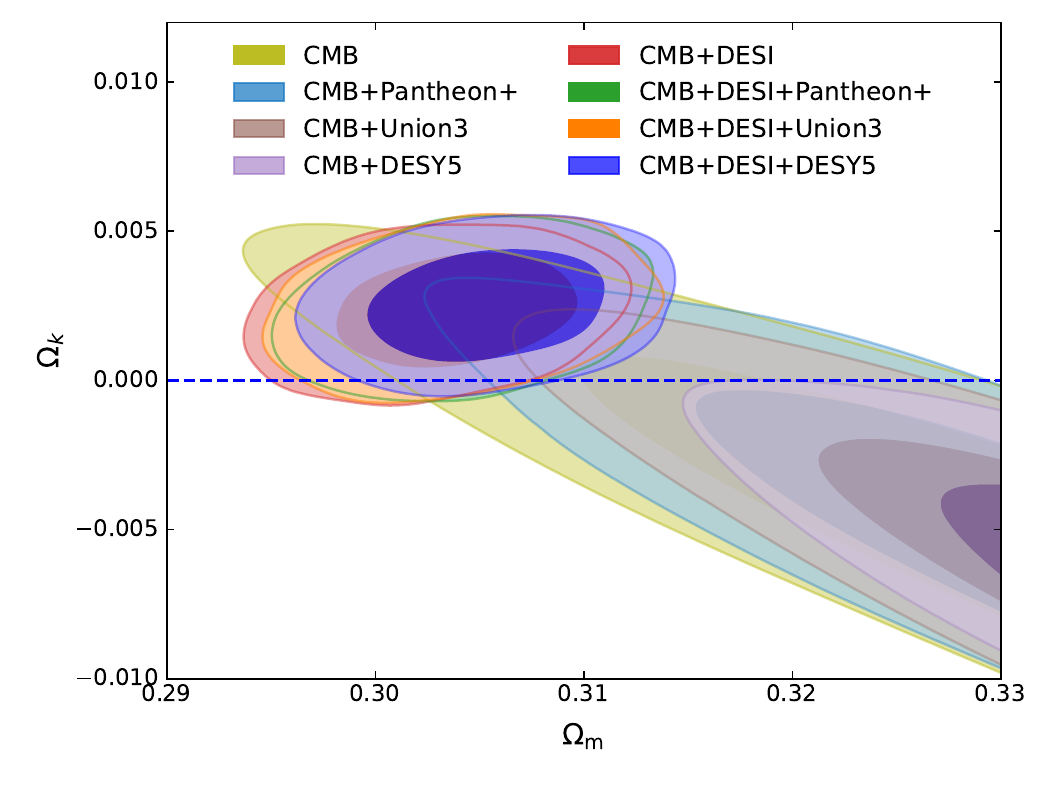}
	\caption{Left panel: one-dimensional probability distributions of the curvature parameter from high-redshift probes such as CMB, Planck  and from each low-redshift probe considered here independently. Right panel: two-dimensional allowed contours at the 68\% and 95\% ~CL in the ($\Omega_k$, $\Omega_m$) plane from CMB and from CMB plus a number of different cosmological low-redshift probes.}\label{fig:omk}
\end{figure*}

Recently, the newly released DESI DR2 BAO measurements provide the constraint $\Omega_k=0.025\pm 0.041$, being consistent with zero curvature at the $1\,\sigma$ level~\cite{DESI:2025zgx}. It is interesting that the combination of recently released DESI DR2 BAO measurements and CMB data gives a positive curvature with $\Omega_k=0.0023\pm 0.0011$~\cite{DESI:2025zgx}, indicating a beyond $2\,\sigma$ evidence for an open universe, which is clearly inconsistent with Planck's predictions. 

This result implies that the addition of DESI DR2 to CMB leads to a transition from $\Omega_k<0$ to $\Omega_k>0$~\cite{Chaudhary:2025pcc, Chaudhary:2025uzr}. As a consequence, a question naturally arises: Is there a curvature tension between the early universe and late universe observations? In this study, we will clarify the current status of curvature constraints and address completely this issue by using the early-time CMB and late-time distance probes such as BAO and SN. 

The structure of the manuscript is as follows. In the following section, we shall describe the cosmological framework. In Sec.~\ref{sec:dataandmethods}, we explain the cosmological observations employed in the data analyses and the methodology used along the manuscript. Section~\ref{sec:results} contains the most important findings of the analysis presented here, to conclude in Sec.~\ref{sec:conclusions}.

\begin{figure*}
	\centering
	\hspace*{-1cm}
	\includegraphics[scale=0.47]{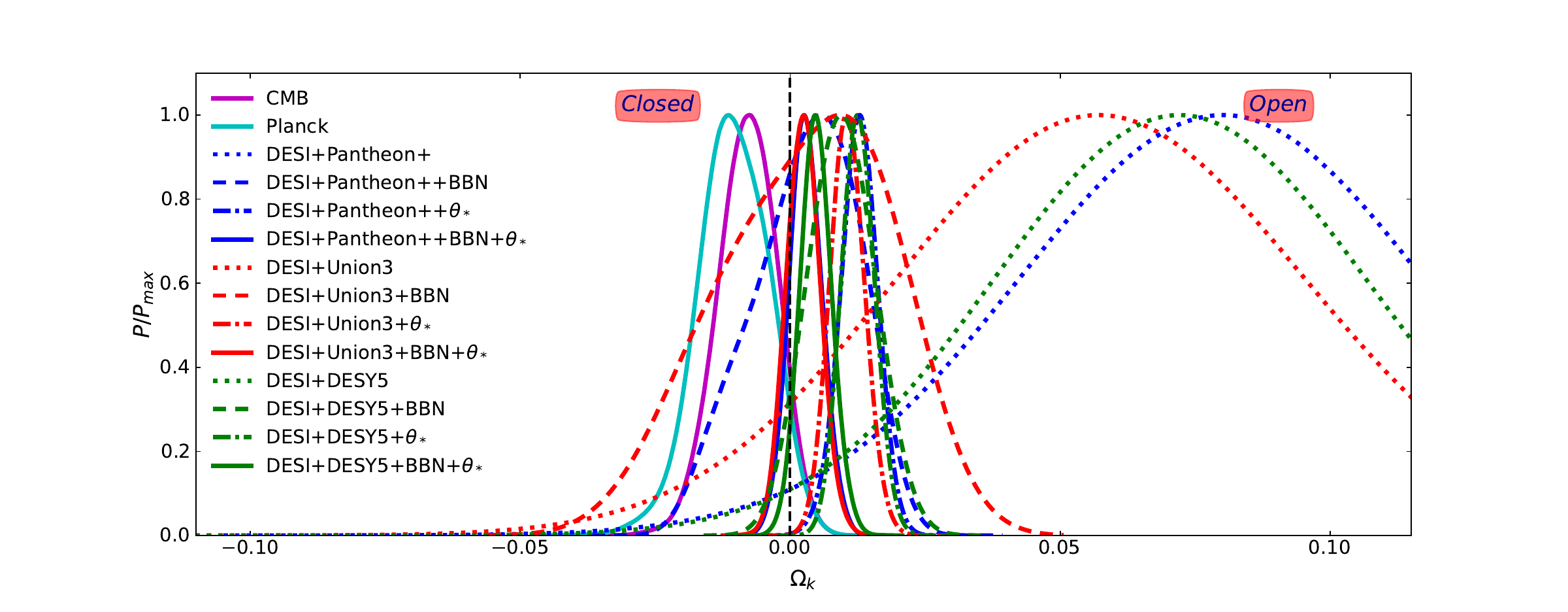}
	\caption{One-dimensional probability distributions for CMB, Planck and DESI, plus the former combined with different Supernovae Ia luminosity distance datasets (Pantheon+, Union3 or DESY5) and/or BBN/angular diameter distance priors.}\label{fig:tension}
\end{figure*}

\section{Basics} 
\label{sec:basics}
In the framework of general relativity~\cite{Einstein:1916vd}, a homogeneous and isotropic universe is characterized by the Friedmann-Lema\^{i}tre-Robertson-Walker metric (FLRW) metric
\begin{equation}
\mathrm{d}s^2=-\mathrm{d}t^2+a^2(t)\left[\frac{\mathrm{d}r^2}{1-Kr^2}+r^2\mathrm{d}\theta^2+r^2\mathrm{sin}^2\theta \mathrm{d}\phi^2\right],      \label{eq:FRW}
\end{equation}  
where $a(t)$ and $K$ are the scale factor at cosmic time $t$ and the Gaussian curvature of spacetime, respectively. The Friedmann equations read as $H^2=(8\pi G\rho)/3$ and $\ddot{a}/a=-4\pi G(\rho+3p)/3$,  where $H$ is the cosmic expansion rate and $\rho$ and $p$ are the mean energy density and pressure of different species including radiation, baryons, DM and DE. For the late universe, we ignore the radiation contribution to the background evolution of the universe. Then, the normalized Hubble parameter $E(a)\equiv H(a)/H_0$ is expressed as 
\begin{equation}
E(a)=\left(\Omega_{m}a^{-3}+\Omega_{k}a^{-2}+1-\Omega_{m}-\Omega_{k}\right)^{\frac{1}{2}}, \label{eq:ezcpl}
\end{equation}
where $\Omega_m$ and $\Omega_{k}\equiv -K/H_0^2$ are today's matter fraction and cosmic curvature, respectively. It reduces to $\Lambda$CDM when $\Omega_k=0$.

\section{Data and Methods} 
\label{sec:dataandmethods}

We use the Planck 2018 high-$\ell$ \texttt{plik} temperature (TT) likelihood at multipoles  $30\leqslant\ell\leqslant2508$, polarization (EE) and their cross-correlation (TE) data at $30\leqslant\ell\leqslant1996$, and the low-$\ell$ TT \texttt{Commander} and \texttt{SimAll} EE likelihoods at $2\leqslant\ell\leqslant29$~\cite{Planck:2019nip}. We adopt conservatively the Planck lensing likelihood from \texttt{SMICA} maps at $8\leqslant\ell \leqslant400$~\cite{Planck:2018lbu} and the CMB lensing maps from the the Atacama cosmology telescope (ACT) DR6 at $40\leqslant\ell \leqslant763$~\cite{ACT:2023dou}.
We take 13 BAO measurements from DESI DR2 including the BGS, LRG1, LRG2, LRG3+ELG1, ELG2, QSO and Ly$\alpha$ samples at effective redshifts $z_{\rm eff}=0.295$,  0.51, 0.706, 0.934, 1.321, 1.484 and $2.33$, respectively~\cite{DESI:2025zpo,DESI:2025zgx,DESI:2025fii}. We use three uncalibrated SN datasets: (i) Pantheon+ consisting of 1701 data points from 18 different surveys in $z\in[0.00122, 2.26137]$~\cite{Brout:2022vxf}; (ii) Union3 with 22 spline-interpolated measurements derived by 2087 SN from 24 different surveys in $z\in[0.05, 2.26]$~\cite{Rubin:2023jdq}; (iii) DESY5 including 1735 effective data points in $z\in[0.025, 1.130]$~\cite{DES:2024jxu}. 
Hereafter, we use ``CMB'' and ``Planck'' to denote the datasets with and without ACT DR6 lensing, respectively.

We take the Boltzmann solver \texttt{CAMB}~\cite{Lewis:1999bs} to calculate the background dynamics and perturbation evolution of the universe. To implement the Bayesian analysis, we employ the Monte Carlo Markov Chain (MCMC) method to infer the posterior distributions of model parameters using \texttt{Cobaya}~\cite{Torrado:2020dgo}. We assess the convergence of MCMC chains by the Gelman-Rubin criterion $R-1\lesssim 0.01$~\cite{Gelman:1992zz} and analyze them using \texttt{Getdist}~\cite{Lewis:2019xzd}. 

We use the uniform priors for model parameters: the baryon fraction $\Omega_bh^2 \in [0.005, 0.1]$, CDM fraction $\Omega_ch^2 \in [0.001, 0.99]$, acoustic angular scale at the recombination epoch $100\theta_* \in [0.5, 10]$, spectral index $n_s \in [0.8, 1.2]$, amplitude of the primordial power spectrum $\ln(10^{10}A_s) \in [2, 4]$, optical depth $\tau \in [0.01, 0.8]$ and $\Omega_{k} \in [-1, 1]$. To completely explore the parameter space preferred by data, we use a wide enough prior of $\Omega_{k}$ and impose the condition $\Omega_{k} + \Omega_{b} + \Omega_{c} \leqslant 1$ in the Bayesian analysis. Note that the $\Omega_{k}$ prior we use is much wider than $\Omega_{k} \in [-0.3, 0.3]$ used by the DESI collaboration~\cite{DESI:2025zgx}. When implementing an alternative comparison, we adopt the Big Bang Nucleosynthesis (BBN) prior $\Omega_bh^2=0.02218 \pm 0.00055$ estimated by~\cite{Schoneberg:2024ifp} based on the \texttt{PRyMordial} package~\cite{Burns:2023sgx} including marginalization over uncertainties in nuclear reaction rates, and also use a conservative ``BAO-only'' prior $100\theta_*=1.04110 \pm 0.00053$~\cite{DESI:2025zgx} from Planck CMB data. 

\begin{table*}[!t]
	\renewcommand\arraystretch{1.6}
	\begin{center}
		\caption{Mean values and $68\%$~CL errors on the curvature, the matter mass energy density and the Hubble constant from CMB, Planck, DESI and combinations of these measurements with additional data sets, such as those coming from SN Ia luminosity distances or priors from BBN and CMB acoustic scale. We also show the tension with CMB ($T_c\,(\sigma)$) and Planck ($T_p\,(\sigma)$) observations, as well as the significance for an open universe ($S(\sigma)$). The stars refer to the fact that $H_0$ is unconstrained, while the diamonds on $S(\sigma)$ are related to the fact that either the data sets are in tension or the data sets imply a closed universe. }
		\setlength{\tabcolsep}{3.6mm}{
			\label{tab:CPL}
			\begin{tabular}{l c c c c c c}
				\hline
				\hline
				Parameter & $\Omega_k$ & $\Omega_m$ & $H_0$ & $T_c\,(\sigma)$ & $T_p\,(\sigma)$  & $S\,(\sigma)$    \\
				\hline 
				CMB &  $-0.0078\pm 0.0055$      & $0.340\pm 0.019$ & $64.6^{+1.8}_{-2.1}$ & $0$ & $0.31$ & $\blacklozenge$\\
				Planck &  $-0.0105\pm 0.0066$   & $0.351\pm 0.024$ & $63.6^{+2.1}_{-2.4}$ & $0.31$ & $0$ & $\blacklozenge$\\
				CMB+DESI &  $0.0022\pm 0.0012$  & $0.3028\pm 0.0038$ & $68.65\pm 0.33$    & $1.78$ & $1.89$ & $\blacklozenge$\\
				CMB+Pantheon+ &  $-0.0070^{+0.0045}_{-0.0041}$  & $0.337\pm 0.014$ & $64.9\pm 1.5$ & $0.12$ & $0.45$ & $\blacklozenge$\\
				CMB+Union3 &  $-0.0094\pm 0.0048$               & $0.346\pm 0.016$ & $64.0\pm 1.6$ & $0.22$ & $0.13$ & $\blacklozenge$\\
				CMB+DESY5 &  $-0.0099\pm 0.0042$                & $0.348\pm 0.014$ & $63.8\pm 1.3$ & $0.30$ & $0.08$ & $\blacklozenge$\\
				CMB+DESI+Pantheon+ &  $0.0024\pm 0.0013$        & $0.3042\pm 0.0037$ & $68.56\pm 0.32$ & $1.80$ & $1.92$ & $\blacklozenge$\\
				CMB+DESI+Union3 &  $0.0024\pm 0.0013$           & $0.3039\pm 0.0038$ & $68.58\pm 0.33$ & $1.80$ & $1.92$ & $\blacklozenge$\\
				CMB+DESI+DESY5 &  $0.0025\pm 0.0012$            & $0.3054\pm 0.0038$ & $68.49\pm 0.32$ & $1.83$ & $1.94$ & $\blacklozenge$\\
				\hline
				DESI+Pantheon+ &  $0.080\pm 0.037$  & $0.287\pm 0.013$ & $\bigstar$ & $2.35$ & $2.41$  & $2.17$\\
				DESI+Union3 &  $0.058\pm 0.038$     & $0.291\pm 0.012$ & $\bigstar$ & $1.71$  & $1.78$ & $1.56$\\
				DESI+DESY5 &  $0.072\pm 0.034$      & $0.290\pm 0.012$ & $\bigstar$ & $2.32$ & $2.38$  & $2.12$\\
				DESI+Pantheon++BBN &  $0.0028^{+0.0120}_{-0.0082}$  & $0.3046\pm 0.0076$ & $68.61^{+0.60}_{-0.68}$ & $1.07$ & $1.26$ & $0.23$\\
				DESI+Union3+BBN &  $0.003^{+0.018}_{-0.015}$        & $0.3040\pm 0.0088$ & $68.54^{+0.81}_{-0.92}$ & $0.68$ & $0.82$ & $0.11$\\
				DESI+DESY5+BBN &  $0.0095\pm 0.0061$                & $0.3088\pm 0.0077$ & $68.32\pm 0.59$         & $2.11$ & $2.23$ & $1.48$\\
				DESI+Pantheon++$\theta_*$ &  $0.0127\pm 0.0034$  & $0.3079^{+0.0094}_{-0.0076}$ & $\bigstar$ & $3.17$ & $3.12$ & $3.33$\\
				DESI+Union3+$\theta_*$ &  $0.0106\pm 0.0032$     & $0.3026\pm 0.0077$ & $\bigstar$ & $2.89$ & $2.88$ & $3.51$\\
				DESI+DESY5+$\theta_*$ &  $0.0126\pm 0.0031$      & $0.3078\pm 0.0071$ & $\bigstar$ & $3.23$ & $3.17$ & $>5$\\
				DESI+Pantheon++BBN+$\theta_*$ &  $0.0029\pm 0.0031$  & $0.3046\pm 0.0076$ & $68.59\pm 0.53$ & $1.69$ & $1.84$ & $0.96$\\
				DESI+Union3+BBN+$\theta_*$ &  $0.0025\pm 0.0031$     & $0.3039\pm 0.0078$ & $68.51\pm 0.54$ & $1.63$ & $1.78$ & $0.77$\\
				DESI+DESY5+BBN+$\theta_*$ &  $0.0048\pm 0.0030$      & $0.3102\pm 0.0073$ & $68.58\pm 0.54$ & $2.01$ & $2.11$ & $1.61$\\
				
				\hline
				\hline
		\end{tabular}}
	\end{center}
\end{table*} 

\section{Results}
\label{sec:results}
Figure \ref{fig:omk} summarizes the main findings of our analyses. The left panel shows the fact that low-redshift data clearly prefers an open universe. That is the case for DESI, Pantheon+, Union3 and DESY5 data. On the other hand, Planck and CMB observations prefer a closed universe with $\Omega_k<0$. This disagreement in the inferred value of the cosmic curvature is also clearly visible from the results of Fig.~\ref{fig:omk}, which depicts the two-dimensional allowed contours at the 68\% and 95\% ~CL in the ($\Omega_k$, $\Omega_m$) plane from CMB and from CMB plus a number of different cosmological low-redshift probes. The key issue here is the fact that high-redshift observations such as CMB or Planck alone prefer a negative curvature, and this very same preference still holds when combining CMB observations with low-redshift probes such as SN Ia luminosity distance probes. This statement is true as long as DESI data are not considered in the data analyses. When these observations are combined with the CMB ones, the preference for a negative curvature disappears and changes to a preference for a positive one, regardless whether or not SN Ia are also present in the different data combinations. 
Notice that the tension between these two measurements of the cosmic curvature, that is, the one from CMB alone and that derived once DESI observations are also considered in the numerical analyses, is closely related to the existing discrepancy between the value of the present matter density extracted from Planck CMB data, $\Omega_m=0.3169\pm 0.0065$, and that extracted from DESI observations, $\Omega_m= 0.2975\pm 0.0086$. The tension in the $\Lambda$CDM framework reaches the $2.3~\sigma$ level~\cite{DESI:2025zgx}. Given the degeneracy between $\Omega_m$ and $\Omega_k$, a larger value of the matter density will lead to the negative curvature region, as illustrated in the right panel of Fig.~\ref{fig:omk}. Therefore, it seems that the value of $\Omega_m$ increases as the observational value of the redshift does. It has been also highlighted the fact that the discrepancy still persists when one replaces the DESI BAO measurements~\cite{DESI:2024uvr,DESI:2024lzq,DESI:2024mwx}  with the 
DESI full-shape galaxy clustering ones~\cite{DESI:2024hhd,DESI:2024jxi}.

Figure~\ref{fig:tension} fully illustrates the tension between low and high-redshift observations concerning curvature constraints. It is clearly visible that CMB and Planck lie in the closed universe region, while DESI together with other combinations lie in the open universe one.
The strongest preference for an open universe arises when DESI is combined with Supernova Ia luminosity distance measurements. The data combinations DESI+Pantheon+, DESI+ Union3 and DESI + DESY5 prefer an open universe with a significance of $\sim 2\sigma$. However, this strong preference
is notably reduced when high-redshift priors, such as BBN and angular diameter priors are also considered in the data analyses. Indeed, for the data combinations DESI+ Pantheon++BBN+$\theta_\star$, DESI+ Union3+BBN+$\theta_\star$ and DESI+ DESY5+BBN+$\theta_\star$ the preference for an open universe almost dilutes. This effect clearly states the fact that while low-redshift data prefer open universes with a positive curvature, high-redshift observations prefer a closed universe with a negative $\Omega_k$ component. 

Table~\ref{tab:CPL} illustrates the mean values of the curvature, the matter density and the Hubble constant, together with the $95\%$~CL errors as well as the tensions with CMB  or Planck observations and the significance for a positive curvature, characterizing an open universe. We show these quantities for a number of possible data sets combinations of CMB or DESI. Notice that we do not report the significance for the combinations of CMB with other data due to the existing tensions between CMB and DESI measurements. 
First of all, let us focus on the combinations with CMB observations in Tab.~\ref{tab:CPL}. When combining CMB with SN Ia data, the constraining power of SN is not enough to shift the negative values of the curvature preferred by Planck and CMB data to positive ones. Once DESI observations are also addressed, the curvature values are shifted to positive ones, and the significance for an open universe rises to the $2\sigma$ level, as well as the tension with either CMB or Planck measurements, that also reaches $2\sigma$. The lower value of $\Omega_m$ preferred by DESI data combinations translates into a larger value of $H_0$, ameliorating the Hubble constant tension.

Table ~\ref{tab:CPL} also shows the combinations of low-redshift data alone and also with either BBN and/or angular diameter distance priors. When considering only DESI pluys Supernova IA luminosity distance measurements, an open universe is preferred at the $2\sigma$ level, and the tension with CMB or Planck observations raises also the very same significance. Among the three possible SN Ia luminosity distance surveys explored here, the Union3 one provides the lowest tension ($\sim 1.7\sigma$), while the Pantheon+ one leads to the largest tension ($\sim 2.4\sigma$. Indeed, the DESI + Pantheon+ combination results in the lowest value of $\Omega_m$. However, some of these results are not significant once a high-redshift prior from BBN is also considered. For instance, for the DESI+Union3+BBN data combination the tension and the preference for an open universe dilutes. Nevertheless, for the case of DESI+DESY5+BBN the preference for a positive curvature $2\sigma$ still remains at the $2\sigma$ level. Such a preference reaches the $5\sigma$ level if the BBN prior is replaced by a prior on the angular diameter distance. These combinations, that is, DESI plus SN Ia luminosity distances plus a prior on $\theta_{\star}$ lead to the strongest preference for open universes, surpassing in almost all the cases the $3\sigma$ significance. Finally, once all the data sets are considered together, the significance for open universes is reduced but there is still a $2\sigma$ tension when considering  the DESY5 SN Ia survey data.

\section{Discussions and conclusions}
\label{sec:conclusions}

Even if the standard cosmological scenario provides an excellent fit to a plethora of observational data sets, in the last years a number of tensions have been reported, being the one on the Hubble constant measurements the most significant one. These tensions appear between high and low-redshift measurements of the very same quantity. While systematical issues may play a role, enlarging the minimal cosmological scenario can also provide an explanation of some of the current tensions. In this manuscript, we notice the existence of another tension, related to the cosmic curvature. Inflationary predictions state that the curvature of the universe must be very close to zero and therefore the universe's geometry is expected to be flat, and any departure from flatness would imply a failure of the (minimal) inflationary models. Therefore, assessing the evidence for a flat universe with $\Omega_k=0$ is a timely study, especially when new cosmological observations become available. In this regard, it is also crucial to confront the predictions from low- and high-redshift observations. 
Here we report a raising tension between low- and high-redshift measurements considering the universe's curvature: while Planck's CMB data prefer a closed universe, integrated observations of DESI and SN Ia reveals an open universe with $\Omega_k>0$ at the $2\sigma$ level. This preference can even reach the $5\sigma$ level when considering DESY5 SN Ia luminosity distance data and a prior on $\theta_{\star}$, challenging inflationary predictions and raising the tension between DESI observations and Planck or CMB ones above $3\sigma$. The reason for this discrepancy is partly that DESY5 SN Ia data is more sensitive than Pantheon+ or Union3 samples to the large compression of the background parameter space when combining early- and late-time observations. It is therefore of utmost importance to assess the existence of this raising curvature tension, and future cosmological observations have the key for that purpose. 

The standard canonical single-field slow-roll inflation strongly drives the curvature towards zero. To end up measurably open after the ordinary slow-roll, a shorter-than-usual inflationary period or the special initial condition is needed. Hence, in general, one would not expect a non-flat universe in this vanilla scenario. While avoiding the graceful-exit problem, models of open inflation where the metastable false vacuum decays to the true vacuum via quantum tunneling can naturally produce an small but nonzero positive curvature. As a consequence, our results may provide possibly observational support for open inflation theories.

\section*{Acknowledgements}
This work has been supported by the Spanish
grant PID2023-148162NB-C22 and by the European SE project ASYMMETRY (HORIZON-MSCA-2021-SE-01/101086085-ASYMMETRY) and well as by the Generalitat Valenciana grants PROMETEO/2019/083 and CIPROM/2022/69. DW is supported by the CDEIGENT Fellowship of Consejo Superior de Investigaciones Científicas (CSIC).
OM acknowledges the financial support from the MCIU with funding from the European Union NextGenerationEU (PRTR-C17.I01) and Generalitat Valenciana (ASFAE/2022/020).
SC acknowledges the support of  Istituto Nazionale di Fisica Nucleare (INFN), Sez.  di Napoli,  {\it Iniziative Specifiche} QGSKY and MOONLIGHT2. 
DFM thanks the Research Council of Norway for their support and the resources provided by UNINETT Sigma2 -- the National Infrastructure for High-Performance Computing and Data Storage. 
This paper is based upon work from COST Action CA21136 {\it Addressing observational tensions in cosmology with systematics  and fundamental physics} (CosmoVerse) supported by COST (European Cooperation in Science and Technology). 

\bibliography{main}

\end{document}